\documentstyle[epsf,twoside,fleqn,espcrc2]{article}

\newcommand{\AmS}{{\protect\the\textfont2
  A\kern-.1667em\lower.5ex\hbox{M}\kern-.125emS}}

\title{Gravity and Random Surfaces on the Lattice: A Review}

\author{D. A. Johnston\address{Mathematics Department, Heriot-Watt University, \\
        Edinburgh, EH14 4AS, United Kingdom}
	}
       
\begin{document}

\begin{abstract}
We review recent work in the lattice approach
to random surfaces and quantum gravity. 
Our task is made somewhat easier by some very interesting
results, particularly in four dimensions, that have
appeared recently and which are reported elsewhere
in these proceedings.
Inevitably, given the scope of the review
and the limitations of space, the
presentation will omit work of importance
and be telegraphic in discussing work that is included,
for which apologies are offered in advance.
After the customary brief historical
introduction we work our way 
in dimensional order from one up to four
dimensions before closing with
some remarks on the relation, if any,  between the various lattice
models and ``real'' 4D gravity.
\end{abstract}

\maketitle

\section{INTRODUCTION}

It is probably fair to say that lattice gauge
theory in the large is now a mature subject.
The theoretical underpinnings are understood
and the numerics are under control. The subject of this
review is a much more speculative application
of lattice methods, namely to the study of quantum
gravity in various dimensions.
Geometry itself becomes dynamical in gravitation,
so it is clear that unlike standard lattice theories
we will be dealing with dynamical lattices of some
sort in which the geometry and the matter living
on the lattice, if any, are in interaction.

The first question to ask before embarking
on the numerical investigation of any theory
is why bother? In the case of gravity,
the answer is quite clear: the Einstein-Hilbert
action for general relativity 
is perturbatively non-renormalizable,
essentially because of the 
dimensionful coupling.
Although this does not exclude a
valid theory
\footnote{An example
of a perturbatively non-renormalizable
theory that does have a well-defined continuum limit
is the $4D$ $O(4)$ non-linear
sigma model \cite{1}.}, it does force us to employ methods
other than those of perturbative
quantum field theory to investigate 
the model. There are further problems
lurking even after the rotation to Euclidean signature
which is generically necessary to obtain tractable
simulations on the lattice. These are due to the
unboundedness below of the Euclidean 
action,
due to conformal mode fluctuations, 
$$
S = \int_{M} d^4 x \sqrt{g} \left(
\Lambda - { 1 \over 16 \pi G} R \right) 
$$
where $M$ is our spacetime manifold.
At first sight 
this might seem to render even the Euclidean
partition function
$$
Z = \int [ D g_{ab} ] \exp \left( - S 
\right),  
$$
ill-defined.  We might be saved, however,
by the measure in the path integral 
giving negligible weights to such configurations.
In order to see if such a happy occurence does take place
we need the framework of some regulated lattice theory
where the integral/sum over configurations can be clearly
defined.

It is perhaps worth strengthening the case
for numerical investigations by noting that 
minimal tinkerings with the quantum field 
theory don't work. Introducing higher derivative
terms gives a renormalizable theory with action
$$
\int_{M} d^4 x \sqrt{g} \left(
\Lambda - { 1 \over 16 \pi G} R  
+  \alpha R_{\mu \nu} R^{\mu \nu}  
+ \beta R^2 
\right) 
$$
which is is even asymptotically free for suitably chosen
$\alpha$ and $\beta$ \cite{2} but, as might be expected,
one runs into problems with unitarity.
More radical solutions, SUSY, SUGRA, strings, p-branes, M-theory,
non-commutative differential geometry have all generated a lot
of beautiful mathematics (and even more preprints)
but getting back to boring old four
dimensions and a non-supersymmetric, or at least
broken supersymmetric particle spectrum, poses problems
of varying degrees for all of them. 

A more modest line of attack is to
attempt to formulate a non-perturbative approach
to the Einstein-Hilbert action and its close relatives.
There are two complementary schemes that have been used
in most recent work, 
Regge Calculus (RC) and Dynamical Triangulations (DT),
which we now describe briefly in turn.

{\bf Regge calculus (RC):}
was originally suggested by Regge in 1961 \cite{3}
as a means of discretizing classical GR.
The recipe for discretizing
the D-dimensional spacetime manifold $M$
was to replace it by D-simplices.
Curvature then resides on the $D-2$ 
dimensional hinges between the simplices
(i.e. vertices in $2D$ gravity, links
in $3D$ gravity and faces in $4D$ gravity).
In the RC approach 
the discretized equivalent of the integration
over metrics
is usually implemented by taking
a {\it fixed} connectivity
simplicial complex with {\it varying}
edge lengths. Regge and Ponzano
came back for a second bite with a
theory of quantum $3D$ gravity in 1968
\cite{4} and more recently there
has been much numerical work on the
quantum theory in $4D$ by Hamber \cite{5}
and the Vienna group \cite{6}.

{\bf Dynamical Triangulations (DT):}
The recent work with this approach is an outgrowth of 
ideas that had their genesis in
the theory of random surfaces and 2D
gravity \cite{7} but 
this variant also has a long pedigree,
in that the basic ideas were formulated 
Weingarten in 1982 \cite{8} using hypercubes.
As with RC, one replaces $M$ by D-simplices,
but now we take {\it varying} connectivity
in our 
simplicial complex and {\it fixed} 
edge lengths in order
to implement a discretized version
of the integral over metrics. 
In two dimensions this formulation is essentially
a recipe for the numerical evaluation
of matrix model
partition functions and has been an
unqualified success. As
a cautionary remark it should be noted that one does not
have the same theoretical underpinning in three
and four dimensions.

From the statistical mechanical viewpoint
in both RC and DT we are dealing with theories
with a particular sort of annealed disorder
coming from the geometric fluctuations.
This is as we might expect - the dynamics
of the lattice should be taking place on
the same timescale as any matter, so we
have annealed rather than quenched averages
to contend with.
In RC we have annealed {\it bond length} disorder
whereas in DT we have annealed {\it connectivity}
disorder.
 
In both RC and DT 
the discretized actions are appealingly simple,
being for Regge Calculus in $4D$
$$ S = \beta \sum_t A_t \delta_t - \lambda \sum_s V_s $$
where $A_t$ is the area of triangular faces $t$,
$V_s$ is the volume of simplices $s$
and the deficit angle is given by
$$ \delta_t = 2 \pi - \sum_{t \epsilon s} \theta_{s,t} $$
and for dynamical triangulations in various 
dimensions
\begin{eqnarray}
S &=& \kappa_4 N_4 - \kappa_2 N_2  \; \; (4D) \nonumber \\
S &=& \kappa_3 N_3 - \kappa_1 N_1 \; \; (3D)  \nonumber \\
S &=& \kappa_2 N_2 \; \; \; \; \; (2D) \nonumber 
\end{eqnarray}
where the $N_k$ are the numbers of $k$-simplices
and the various $\kappa$'s are the coupling constants.

There are various
choices to be made in the two schemes.
In RC the measure
for the integration over the edge lengths
must be chosen according to some
prescription and a fixed triangulation decided upon.
In DT one must decide which class of simplicial
complexes to employ - for instance in $2D$
the finite size effects are minimized, rather
counterintuitively, by allowing various
degenerate triangulations.
In both schemes one must decide which topology
to fix or whether, indeed, to let it vary.

In DT the moves that change the connectivity
in $D$ dimensions can be obtained from 
``slicing'' $D+1$ dimensional simplices
appropriately. In two
dimensions there is a ``flip''
move that preserves the number of vertices
in a triangulation
as well as an insert/delete move, so it is
possible to cary out either
{\it canonical} fixed number of nodes simulations
or
{\it grand canonical}
varying number of nodes simulations.
The moves in two dimensions are shown overleaf:

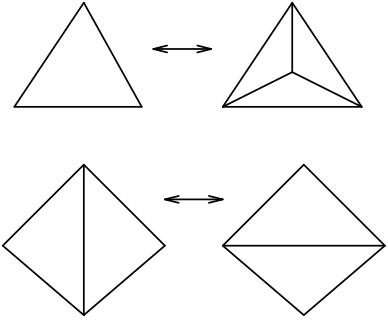
\begin{figure}[h]
\centerline{ \epsfxsize=2in
             \epsffile{fig1.eps} }
\end{figure}

\bigskip

This option is no longer open
in three and four dimensions 
where the moves 
change the number of simplices.
These moves are represented
in their simplest form
as so-called $(k,l)$ moves
\cite{9}
which take $k$ subsimplices
to $l$ subsimplices,
and may be shown to be ergodic.
The moves in the three
dimensional case are shown below:

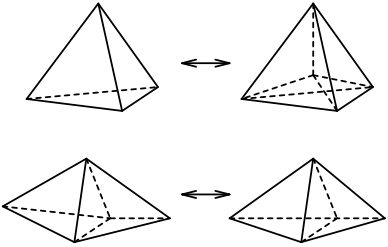
\begin{figure}[h]
\centerline{ \epsfxsize=2in
             \epsffile{fig2.eps} }
\end{figure}

\noindent
being $(1,4)$ and $(2,3)$
moves respectively. My
very limited skills as an
illustrator defeat me in the four dimensional
case where the $(k,l)$ moves are $(1,5)$,
$(2,4)$ and $(3,3)$.
In addition cluster-like moves
have also been defined, which have the effect
of cutting off large chunks of the 
simplicial complex and gluing them back on elsewhere.
 
The partition function of 
interest in the $4D$ simulations
carried out using the moves
described above is
$$
Z(\kappa_2, \kappa_4) = \sum_{N_2, N_4} Z_{N_2,N_4} \exp \left( - S \right)$$
where
$$Z_{N_2,N_4} = \sum_{T(N_2,N_4)} W(T)$$ 
with $W(T)$ being the symmetry factor for a simplicial complex
and $S$ being the $4D$ DT action.
To discuss possible
critical behaviour it is
convenient to define a fixed volume partition function
$$
Z(N_4, \kappa_2) = \sum_{T(N_4)} \exp \left( \kappa_2 N_2 \right)$$
so
$$Z(\kappa_2,\kappa_4) = \sum_{N_4} Z(N_4, \kappa_2) \exp \left( - \kappa_4 N_4
\right)$$
As we have noted it is impossible 
to carry out a fixed volume simulation in $4D$
because an ergodic set of $(k,l)$ moves entails changes
in the number of simplices, so
we perform a 
fudge to stay ``close'' to given $N_4$
by modifying the action with
a gaussian term
$$ S = \kappa_4 N_4 - \kappa_2 N_2 
+ \gamma ( N_4 - V)^2 $$

The observables in 
our simulations are 
the discretized versions of
curvature and the associated susceptibilities
$$<R> \simeq \left<  N_2 \over N_4 \right>$$
$$\chi_R \simeq \left< N_4 \right> \left( <R^2> - <R>^2 \right)$$
as well as other
geometrical properties such mean 
geodesic distances, hausdorff dimensions
and curvature correlators.

A schematic phase diagram
for the theory is shown below,
\begin{figure}[h]
\centerline{ \epsfxsize=2in
             \epsffile{fig3.eps} }
\end{figure}
\noindent
where
moving along the bold line representing
the infinite volume limit we may reach a 
critical point at some value of $\kappa_2$
at which correlation lengths may diverge and we can hope to
take a continuum limit.

Extensive simulations by various groups both in DT \cite{10} 
and RC \cite{5,6} over the past few years
suggested the following picture:

\begin{itemize}

\item{} In DT one sees a low $\kappa_2$ ``crumpled'' phase
and a large $\kappa_2$ branched polymer like phase
in both $3D$ and $4D$.

\item{} A fractal baby universe structure was
manifest in $3D$ and $4D$ (and also in $2D$) with DT.

\item{} With DT in $3D$ there was a first order transition
between the crumpled and branched polymer phases.

\item{} With DT in $4D$ the transition was second order.

\item{} RC saw a similar phase structure (but not
necessarily order of transitions).

\item{} Issues of an exponential bound  on 
the number of configurations \cite{12} and
computability in DT appeared to have been settled
to (almost!) everyone's satisfaction.

\end{itemize}

\noindent
The schema above for DT has the great merit of conceptual simplicity
- starting with a very simple action in both $3D$ and $4D$
one has found no continuum limit in $3D$ (fine, there is no graviton
in $3D$) but a continuum limit in $4D$ (also fine, we want to get
a graviton here). It has, unfortunately, the great demerit
of apparently being wrong, as the latest numerical work 
presented in these proceedings indicates.

In the rest of this article we will review the recent work in the field
in one,two, three and four dimensions. The lower dimensional
work is included not just for completeness, a lot of it has a direct
bearing on the understanding of the $4D$ theory. The $1D$
work on polymers, of course, stands on its own merits 
as does the $2D$ work on random surfaces which 
have a wide application outside
the rather esoteric world of quantum gravity simulations and string theory.

\section{1D STRUCTURES AND POLYMERS}

It seems to be a general principle that models
of dynamical geometry in higher dimensions
will, given half a chance, collapse to lower
dimensional structures - the large $\kappa_2$
branched polymer like phase in $3D$ and $4D$
DT being a prime example. It is therefore
of some interest to study simplified models
of branched polymer like objects in their own right.
Indeed, it has been suggested that in $4D$
DT the entire  branched polymer phase may be critical
because of power law correlations observed there \cite{16}.
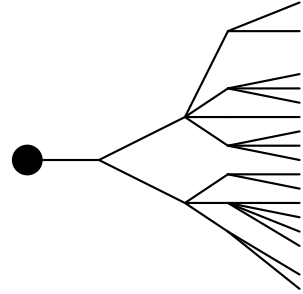
\begin{figure}[h]
\centerline{ \epsfxsize=1.5in
             \epsffile{fig0.eps} }
\end{figure}

\noindent
A note of caution on these observations has been
sounded recently by Bialas \cite{17} who considered
a model of planar rooted planted trees
as shown above.
The partition function for these is just
$$ Z = \sum_{Trees} \rho(T)$$
where the weight, $\rho(T)$, of a given tree depends
on how many vertices $n_k$ of order $k$
are present
$$\rho(T) = t_0^{n_0} \; t_1^{n_1} \; t_2^{n_2} \ldots t_k^{n_k} \ldots$$
with the $t_k$'s being the weights for order $k$ vertices. 

He considered correlations of the form
$$G ( \mu, t, r) = \sum_T \exp ( - \mu n) \rho (T)$$
where $\mu$ is the ``cosmological constant''
with a marked  point a distance $r+1$ from the root,
more
specifically Root-Tail correlations 
$$ \tilde G(\mu,t;r)=\sum_{T_r}d(v_1)d(v_{r+1})e^{-\mu n}\rho(T)$$
where the $d(v)$ are the degrees of the vertices in question.
He found, with some subtleties
in the definitions of correlators, 
$\tilde G = 0$. However, in a canonical (fixed number of vertices) ensemble
there were negative $\sim 1 / r^2$ correlations due to the absence
of short trees. The moral is that the almost fixed volume  
constraint in $4D$ gravity simulations might be responsible
for the power law behaviour seen in the branched polymer phase rather
than being a sign of true critical behaviour. The issue awaits further
clarification.

Planar rooted planted trees were
also used by Ambj\o rn et.al. in an
analytical implementation of fractal blocking \cite{18}.
Various MCRG schemes have been proposed 
for two and higher
dimensional DT simulations. The new feature here, compared
with MCRG on a regular lattice, is that the geometry
is both irregular and dynamical so the choice of basic
blocking move for the lattice itself is not immediately apparent.
One possibility is the so-called fractal blocking or baby universe
renormalization group which makes use of the fractal structure
of the DT simplicial manifolds \cite{20,20a}. In all dimensions these
contain regions where the lattice necks down, which can be 
visualized as baby universes budding off a mother universe.
This can be used to define a blocking move as shown above right
- extremal baby universes (those with no babies themselves)
are lopped off and the couplings rescaled.
\begin{figure}[ht]
\centerline{ \epsfxsize=2.0in
             \epsffile{fig6.eps} }
\end{figure}

The simplicity of the tree model allows an analytical
implementation of this program and shows,
surprisingly that
fractal blocking moves you away from 
a branched structure directly to a linear chain.
Whether this signifies a possible pathology
of the blocking scheme or is merely
a quirk of the model is not entirely clear.
The fact that fractal blocking gives good results
numerically in higher dimensions argues for the latter.

The last result I want to mention in $1D$ concerns
planar rooted trees \cite{21}
where the restriction on having only one branch
emerging from the root node is relaxed. This technical change
gives a soluble model
$$ Z = \sum_T \exp \left( - \mu n (T) \right) \exp \left( - \beta E (T) \right)$$
where
$n(T)$ is the number of vertices, $E(T) = \sum_v \ln k ( v )$
and $k(v)$ is the vertex degree.
There are two phases: a
large $\beta$  branched polymer phase
with lots of short ``bushes'' ;
and a small $\beta$ elongated branched polymer
phase. 
The transition between these is fourth order
$$\phi - \phi_0 = ( \mu - \mu_0)^{1 - \gamma_{string}}$$
where the free energy is 
$\phi = - \ln Z$,  and $\gamma_{string}= 0.3237...$
The result is interesting because it provides
a counterexample to the belief 
that only values of $\gamma_{string}= 1/ n$
for $n=2,3,4,\ldots$ were possible in branched
polymer models. Indeed, by tinkering with the 
vertex weights a range of $\gamma_{string}$
values can be extracted from the model.

\section{2D RANDOM SURFACES}

We make an extended stop in $2D$ also as the DT
approach has its genesis here.
The usual model of interest is the 
Polyakov String 
$$ S = \int d^2 \sigma \sqrt{g} g^{ab} \partial_a \vec X \partial_b 
\vec X
+ \lambda \int d^2 \sigma K^2  $$
where $K^2$ is the extrinsic curvature squared
and acts as a stiffness or rigidity term.
This discretizes neatly to
$$ S = {1 \over 2} \sum_{<ij>} (\vec X^i - \vec X^j)^2
+ \lambda \sum_{\Delta} ( 1 - \vec n_{\Delta_i} \cdot \vec n_{\Delta_j})
$$
where the $n_{\Delta}$ are normals to triangles.
On a dynamical triangulation this action can serve as a
model for a fluid surface,
and on fixed lattice 
(so there is no gravity
at all) as a model for a polymerized or crystalline surface.
Analytically the fluid model appears to be always in a rough phase
but simulations provide evidence for a
crumpling transition as $\lambda$ is increased. 

More recently an alternative action
also based on geometrical ideas,
the
gonihedric string, has been proposed by Savvidy et.al. \cite{26}
$$ S = {1 \over 2} \sum_{<ij>} | \vec X^i - \vec X^j|
\;  \theta(\alpha_{ij})
$$
where
$\theta(\alpha_{ij}) = | \pi - \alpha_{ij} |^{\zeta}$,
$\zeta<1$ and the $\alpha_{ij}$ are the angles
between neighbouring triangles.

Let us start by discussing some recent results 
concerning the MCRG on crystalline surfaces.
There has been some confusion in the past
concerning the nature of the phase diagram and the values
of the critical exponents for the crumpling transition
on crystalline surfaces (which it is
generally agreed {\it does} exist).
Espriu et.al. \cite{27} have developed a
Fourier accelerated Langevin equation approach to the MCRG
on crystalline surfaces
$$\phi(p, t_{n+1}) = \phi(p,t_n) - \Delta t \epsilon(p)
{\cal F} {\partial S \over \partial \phi ( y , t_n)} $$
$$ + \sqrt{ \Delta t \epsilon ( p )} \eta ( p , t_n)$$
$$ \epsilon ( p)  = { max_p \{ \Delta ( \Delta + m^2 )\} \over \Delta ( \Delta + m^2 )}$$
where ${\cal F}$ denotes a Fourier transform.
They blocked 9 geometric operators and measured the
coupling constant flow, finding the exponent $\eta \simeq 1.527(36)$
in agreement with direct simulations, but with a lot less effort.
In addition, qualitative examination of the $\beta$-function gave
support to the presence of a crumpling transition.

The aim of showing conclusively that a flat phase
does exist in the crystalline model (and hence
a crumpling transition) was pursued by the
Syracuse group \cite{28}.
The answer is not immediately obvious because
considering an action of the form
$$S = \sum_{<ij>} \left( |\vec X_i - \vec X_j | - a \right)^2$$
where the $a$ represents an intrinsic length for each edge
gives an effective action for the stress-strain tensor $u_{\alpha \beta}$
$$S \simeq \int d^2 \sigma \left( {\sqrt{3} \over 2}
a^2 u_{\alpha \beta} u_{\beta \alpha} + {\sqrt{3} \over 4} a^2 u^2_{\gamma \gamma} \right)$$
in which the
elastic constants apparently vanish as $a \rightarrow 0$
\footnote{Although their fluctuations diverge.},
so it is difficult to see how a flat phase might be stabilized.

However, measurement of the various scaling exponents
for the running bending stiffness ($\kappa$ in the notation
of \cite{28}, $\lambda$ elsewhere in this review!)
and the running elastic ``constants'', $\mu$ and $\lambda$
$$\kappa(q) \sim q^{-\eta}, \; \; \mu(q) \sim \lambda(q) \sim q^{\eta_u}$$
as well as the roughness exponent $\zeta$
$$<h^2> \sim L^{2 \zeta}$$ 
all of which are related by scaling relations $\zeta = 1 - {\eta \over 2}$,
$\eta = 1 - {\eta_u \over 2}$
gave
$\eta_u = 0.50(1)$ (so from
the scaling relation $\eta = 0.75$),  $\zeta = 0.64(2)$ (thus $\eta = 0.72$
from the other scaling relation).
These consistent estimates
of a positive $\eta$ confirm the existence of a flat phase.

Having surreptitiously slipped in a discussion of surfaces without gravity,
we now return to our remit to look at work
on dynamically triangulated surfaces - in other words, matter
coupled to two dimensional gravity.
One of the most interesting recent developments 
has been the formulation of 
MCRG methods for dynamical lattices. The fractal blocking scheme
has already been described, but another 
scheme, node decimation, has been pioneered by the Syracuse group
\cite{29}. The idea is again simple: we pick a node,
remove it, and then retriangulate the gap to make sure
that we still have a triangulation as shown overleaf.
\begin{figure}[h]
\centerline{ \epsfxsize=2.5in
             \epsffile{fig5.eps} }
\end{figure}

In practice it is simpler to pick a node at random, flip
the surrounding links until it has only three neighbours
and then remove it.
Work with this method has concentrated on 2D simulations,
looking at the flow of geometric operators such as
$${ 1 \over N} \sum_i (q_i - 6)^2, \; \; { 1 \over N} \sum_{<ij>}
(q_i - 6) (q_j - 6)$$
where $q_i$ is the number of neighbours of site $i$,
when a measure factor $\alpha \sum_i \log q_i$
is varied.
The Ising model 
$$S = \beta \sum_{<ij>} \sigma_i \sigma_j$$
coupled to 2D gravity
has also been investigated
and measurements of $\gamma_{string}$
$$Z(N) \sim N^{\gamma_{string}-3} \exp \left( \mu_c N \right)$$
after blocking moves confirm a flow to the Ising value 
$\gamma_{string} = -1/3$.

The other main thrust of recent work in 2D gravity
has been the clarification of scaling
behaviour. This was discussed in some detail by Simon Catterall
in Lattice95 \cite{31}, so I will be very brief here. 
The canonical object of interest in
such measurements for the geometric
sector of the theory is the distribution of
geodesic distances,
given in a grand canonical ensemble with
cosmological constant $\mu$ by
$$G_{\mu} (R)  = \left< \int \int  \sqrt{g} 
\sqrt{g} \delta ( d_g (\xi, \xi') - R ) \right>$$
which can be shown to scale as
$$G_{\mu} (R) \sim \mu^{\nu - \gamma_{string}} F(\mu^{\nu} R), \; \; \nu = 1 / d_H$$
where $d_H$ is the hausdorff dimension of the triangulation
and $F$ is some scaling function.
The other scaling exponent $\eta$
is related to $\gamma_{string}$ by $\gamma_{string} = \nu ( 2 - \eta )$.
The number of points in a radial shell at distance $R$ 
in a microcanonical ensemble of volume $V$ can be
shown to scale as
$$\left< S(R) \right>_V = R^{d_H-1} F \left( { R \over V^\nu} \right)$$
and is perhaps a more amenable observable to measure. 

Simulations \cite{32} have shown that $d_H = 4$ 
when the central charge $c$ of the matter living on the 
triangulation is less than one, so rather
surprisingly the 
back reaction of the matter
does not affect $d_H$. As the triangulations
degenerate to branched polymers for large $c$ it is
also known that $d_H = 2$ in this regime.

What was unclear until very recently was whether
a diverging matter correlation length could be
successfully defined on dynamical triangulations.
Indeed, a toy model of 
spins on a restricted class
of triangulations \cite{33} displayed a transition but no such divergence.
However, simulations presented at this conference \cite{34}
have shown that such a diverging correlation length {\it does}
exist in the Ising and 3-state Potts models on 
the full dynamical triangulations
and also measured the effect of the matter back reaction on the
scaling functions - a delicate task as it is the nose and tail
of the distributions where the effects are seen.

There have also been simulations
\cite{35}
of variants of the standard action
including
$R^2$ terms on dynamical triangulations
with actions of the form
$$S = \beta \sum_i {(6 -q_i)^2 \over q_i} +
\alpha \sum_{<ij>} { ( 6 - q_i ) ( 6 - q_j) \over q_i q_j}$$
A recent analytical solution of dually weighted matrix models
\cite{36} has shown that 
$R^2$ gravity is in the same universality class
as ordinary gravity, which is what is seen in these simulations.
This is rather similar to particles paths weighted by curvature,
which behave as random walks unless the curvature coupling is taken to infinity.

One enduring lacuna in our understanding of 2D gravity is the nature
of the so-called $c=1$ barrier. Matrix models and continuum methods both
fail to get beyond $c=1$, but simulations see no immediate evidence
of pathologies for $c$ values not much larger than one. It is only at large
$c$ that triangulations collapse to branched polymers.
One simulation that casts some light on this problem
appeals to methods from electrical engineering to probe
the nature of the random surface \cite{38}.
One treats a triangulation of spherical topology as a random network
of unit resistors and puts current in and out at two vertices,
measuring the voltage drop across another two. The voltage-current
relation is
$$V(z_1) - V(z_2) = R(z_1,z_2;z_3,z_4) \; I$$
where the resistance is
$$R = - { r \over 2 \pi} \ln | [z_1,z_2,z_3,z_4]|$$
$r$ is the resistivity constant and
$[z_1,z_2,z_3,z_4]$
is the cross-ratio, which appears because of the conformal invariance.
If the theory is to have a sensible continuum limit the distribution of
measured $r$ values should become sharper as the lattice size is increased,
which happens for $c \le 1$ but not, apparently, for $c>1$. This suggests
that the conformal structure of the surface is breaking down at $c=1$.
It is also possible to extract the distribution of modular parameters
on a toroidal triangulation, but finite size effects are rather
stronger here, so the evidence for something nasty happening
just above $c=1$ is weaker.

I have so far relegated RC to a few cursory sentences, which I now
rectify. Some older simulations of the Ising model on a 2D RC lattice
gave an unpleasant surprise \cite{39},
finding results consistent with the Onsager (flat 2D lattice) exponents

\begin{center}
\begin{tabular}{|c|c|c|c|c|c|c|c|} \hline
$q$& $c$          & $\alpha$     & $\beta$       & $\gamma$  & $\delta$& $\nu$  & $\eta$ \\[.05in]
\hline
$2$& $\frac{1}{2}$& $0$          & $\frac{1}{8}$ & $\frac{7}{4}$ & $15$& $1$          & $\frac{1}{4}$ \\[.05in]
\hline
\end{tabular}
\end{center}

\noindent
rather than the expected 
KPZ/DDK exponents for the Ising model coupled to 2D gravity. 
If they were to be believed
RC was failing to simulate gravity in 2D.
Simulations of the Ising and Potts models with DT
{\it do}
find the correct KPZ/DDK exponents.

\begin{center}
\begin{tabular}{|c|c|c|c|c|c|c|c|} \hline
$q$& $c$          & $\alpha$      & $\beta$      & $\gamma$     & $\delta$& $\nu$    & $\eta$ \\[.05in]
\hline
$2$& $\frac{1}{2}$& $-1$          & $\frac{1}{2}$& $2$          & $5$& $\frac{3}{d_H}$ & $2-\frac{2d_H}{3}$\\[.05in]
\hline
\end{tabular}
\end{center}

\noindent
An obvious first reaction is that 
there is a problem with the measure for integrating
over the edge lengths in RC, as this is not prescribed at the outset.
In general a local measure is used, one common choice
being
$$  \prod_{ij} \int { d l^2_{ij} \over l_{ij}^{2 \sigma }}$$
for various $\sigma$. It has been pointed out that if
one thinks of the lattice model as being a discretization of
an exact diffeomorphism invariant
theory then gauge fixing and the resulting Fadeev-Popov determinant
are obligatory \cite{41}. 
The resulting non-local action would be very difficult to simulate 
even in 2D and things get worse
in higher dimensions.
An alternative point of view is that
exact diffeomorphism invariance is recovered in the continuum limit
even when broken on the lattice, 
so we
{\it can} get away with local measure \cite{42}.

There have been a series of very careful investigations
by Janke and Holm of both pure gravity and gravity plus matter
in 2D RC with local measures \cite{43}.
In the pure gravity case they looked at an $R^2$ action 
$$S = \int d^2 \sigma \sqrt{g} \left( \lambda + {a \over 4 } R^2 \right), \; \hat A = { A \over a}$$
which has two scaling regimes
$$\hat A << 1$$
$$Z(A) \sim A^{\gamma_{st}' -3} \exp \left( - { S_c \over \hat A} \right) \exp \left( -\lambda_R A - \xi \hat A \right) $$
giving
$\gamma_{st}' = 2 - 2 ( 1 - g), \; S_c = 16 \pi^2 (1 - g)^2$
and the standard 2D gravity regime
$$\hat A >> 1$$
$$Z(A) \sim A^{\gamma_{st} -3} \exp ( -\lambda_R A )$$
where $\gamma_{st} = 2 - 5/2 ( 1 - g)$.
They found that the case for the failure of RC was
not proven within the limits of accuracy of the simulation,
contrary to some earlier claims.

They also considered Ising matter with a $dl / l$ measure
and found that the exponents
were definitely Onsager. 
It was found, for instance, from the finite size scaling
of the susceptibility maxima that
$${\gamma \over \nu} = 1.745(6) $$
so we can say with some certainty that 2D RC fails
to reproduce gravitational effects
when matter is coupled in. Whether this is simply
a failure of the scheme used in the coupling rather than
the RC itself remains to be seen.

The discussion
of the RC results leads on to
the broader question of what constitutes the universality class of 2D gravity.
There have been various results for Ising and
Potts model simulations that have a bearing on this:

\begin{itemize}

\item{} \cite{45}
Ising spins living in a flat geometry
but with fluctuating connectivity,
in effect a mixture of DT and RC approaches
\footnote{The idea of a mixed
DT/RC approach has also been suggested by Shamir
in the 4D context and investigated analytically \cite{45a}.}, give KPZ/DDK exponents.

\item{} \cite{47}
Ising spins on quenched ensembles of Poisonnian 
random lattices give Onsager exponents.

\item{} \cite{48}
Spins on quenched ensembles of 2D gravity graphs may 
give ``quenched'' KPZ/DDK exponents.

\item{} \cite{49}
Squeezing 2D gravity (truncating the neighbour distribution
in a triangulation to have just $5,6,7$ neighbours)
does not affect the KPZ/DDK values of the exponents. 

\item{} \cite{50}
Spins on any, not just planar, $\phi^3$ graphs 
(the duals of triangulations) give mean field
exponents.

\end{itemize}

\noindent
In the above list \cite{45} appears to show that
it is fluctuating connectivity that is important
rather than curvature fluctuations
in obtaining the KPZ/DDK exponents, whereas \cite{48}
suggests that even with a quenched ensemble of graphs,
and hence no fluctuating connectivities, gravitational effects are still
present. On the other hand, {\it any} collection of random graphs
does not guarantee gravitational effects: Poissonian random graphs
give flat space \cite{47} exponents and generic $\phi^3$ graphs give
mean field exponents \cite{50}. 

Nonetheless,
the vertex number distribution of 2D DT triangulations
can be quite brutally truncated and the effects of gravity 
still observed \cite{49}. Finally, the Onsager
exponents seen in Regge calculus suggest that the fractal structure
of the 2D DT lattices (apparently absent in RC,
but still present in \cite{48,49})
may be an important factor in determining the universality
class.
Exactly what determines the universality class of the 2D gravity
exponents is thus still not clear. 

We close this section with a quick word on plaquette surface
simulations. Such models are
closer to original Nambu-Goto idea 
in which the target space is discretized
$$S = \int d^2 \sigma \sqrt{ \det | \partial_a \vec X \partial_b \vec X | }$$
so strictly speaking gravity has again disappeared from the picture.
For a 2D surface embedded in 3D the general action
$S = \beta_s Area + \beta_l Intersections + \beta_c bends$
maye be written as an Ising like model
$$S_{ising} = J_1 \sum_{<ij>} \sigma_i \sigma_j + J_2 \sum_{<<ij>>} \sigma_i \sigma_j$$
$$+ J_3 \sum_{[i,j,k,l]} \sigma_i \sigma_j \sigma_k \sigma_l$$
where the Ising couplings are related directly to the surface
couplings
$J_1 = {\beta_s + \beta_l \over 2} + \beta_c, \; J_2 = - {\beta_l \over 8} - {\beta_c \over 4}$,
$J_3 = - {\beta_l \over 8} + {\beta_c \over 4}$.
Following this approach for the 
Savvidy gonihedric string action of \cite{26} gives an Ising
model with interesting properties,
including an unusual semi-global symmetry
and Onsager-like exponents \cite{51}.
It has been suggested very recently \cite{52} that the
gonihedric ideas that went into this string action may
also be of relevance for 3D and 4D DT models,
which we now move on to discuss.

\section{HIGHER D}

Some features
of discretized 3D and 4D gravity in the DT approach appear
rather similar to 2D. There is a baby universe
fractal structure and it seems that various scaling
distributions can be defined \cite{53} in both the crumpled
and branched polymer like phases.
One feature of the 4D model 
not shared by its 2D counterpart was
pointed out in \cite{54}:
The vertex order distribution in the crumpled phase 
is very strange.
The distribution $\rho (o)$ of the vertex orders $o(v)$,
display a rather singular behaviour.
\bigskip
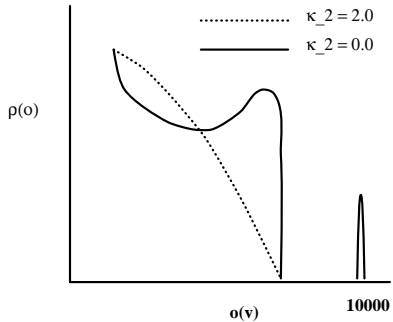
\begin{figure}[h]
\centerline{ \epsfxsize=2.0in
             \epsffile{fig11.eps} }
\end{figure}

In the branched polymer phase at $\kappa_2=2.0$
there is a smooth distribution, as one might have
naively expected, but the crumpled phase
at $\kappa_2=0.0$
has a detached ``spike'' at very large vertex order.
Extensive simulations showed that
this
was {\it not} a finite size effect,
it was
{\it not} a thermalization problem
and that it also appeared for other topologies.

These observations were elaborated on
in \cite{55} where it was
demonstrated numerically that
the generic triangulation of the $D$ sphere for $D>3$ contains one 
singular $D-3$ simplex ($\kappa_2 =0$ for
simplicity, but this is true throughout the crumpled phase).
This singular structure is attached to an extensive
fraction of the rest of the triangulation.
The geometric picture is shown above right for
4D and 5D respectively.
\begin{figure}[h]
\centerline{ \epsfxsize=2.0in
             \epsffile{fig12.eps} }
\end{figure}

In general one finds that the 
local volume associated with
the singular $D-3$ simplex grows as $\sim V^{2 / 3}$
whereas the local volumes associated with the
secondary simplices attached to this grow as $\sim V$.
Some counting arguments were presented in \cite{55}
as to why this should be so.
It is surprising, and perhaps rather disturbing, that
such singular features are present in the theory,
especially given that the observed phase
transition in 4D is apparently tied up with
their appearance and disappearance.

The nature of the 4D transition in DT
has also been the focus of further
work using the fractal blocking MCRG scheme.
As in 2D it is the distribution of geodesic distances
that is of interest, which in discretized form is
$$<r> = \left< {1 \over N^2} \sum_{ij} r_{ij} \right>$$
In 4D, setting $\kappa_2=\kappa$ and $N_4=N$
for conciseness, one has to take account of the running of the
coupling under the blocking move 
$$\delta r = r_N \delta \ln N + r_{\kappa} \delta \kappa$$
so with the physical volume
$N a^4$ fixed, with $a$ the
lattice spacing, we have $\delta \ln {1 \over a} = {1 \over 4} \delta \ln N$.
We can thus deduce
$\delta \kappa$ from measurements
and extract the $\beta$ function
$$ \beta ( \kappa )  \sim {\delta \kappa \over \delta \ln {1 \over a}}$$
The picture that emerged from doing this was of
an ultraviolet fixed point \cite{20a}, not an infrared one as had been
suggested by some continuum treatments.
One puzzling feature of the simulations in \cite{20a} was that the
slope of the $\beta$ function at the fixed point did not appear to be consistent
with a second order transition, which was the generally accepted picture
for DT in 4D. Further work 
looking at the scaling of cumulants such as
$$c_2 (N_4) = { 1 \over N_4} \left[ <N_0^2> - <N_0>^2 \right]$$
where $N_0 = {N_2 \over 2} - N_4 +2$
was carried out to check this \cite{57}.
The finite size scaling form for this is
$$c_2(N_4) \sim N_4^b f\left( ( \kappa_2 - \kappa_2^*) N_4^c \right)$$
where
$b=c=1$ for a first order transition and take non-trivial
values for a second order transition. 
The simulations found the first order values $b=c=1$.
Further evidence for the first order nature of the transition
was found by deBakker \cite{59} who saw
time series flips characteristic of first order transitions,
as well as a double peak histogram structure emerging on
larger lattices.

Simulations discussed at this conference \cite{57}
show that the first order transition is 
not an artifact stemming from the inclusion
of the $\gamma ( N_4 - V)^2$ term in the action -
relaxing the constraint to allow huge volume excursions
has no effect, so we are left with the task of
deciding whether the model should be patched up
to give a second order transition, is still useful
as it stands, or should be consigned to the bin.

The optimists argue that
some things {\it are} right
and in any case there are lattice
theories with first order transitions
and a Coulomb phase for an entire range of couplings.
The power law correlations in the branched
polymer phase might be argued to be just such
a case \cite{61} \footnote{We should
remember the caveat of the polymer example \cite{17}, however.}.  
The DT model also appears to display
gravitational binding
\footnote{$4D$ Regge Calculus would also
appear to be attractive \cite{5}.}, as defining
one and two particle connected propagators  
for scalar test particles shows \cite{62}.

If we insist on getting a continuous transition, various alternatives
suggest themselves. Ray Renken presented evidence \cite{63} using a
generalization of the node decimation scheme to higher dimensions
that suppressing vertices of high order 
with a measure term could weaken the 3D DT transition. A similar effect
may also appear in 4D. 
It might also be possible to use the ideas of \cite{52}
stemming from the gonihedric string, where new
integral invariants for simplicial complexes
corresponding to higher derivative terms were
derived. These would have the effect of 
``stiffening'' the lattices and possibly
changing the nature of the transition.
One might, of course, lose the transition
completely with either of these approaches.

If we are willing to overlook the problems
experienced in 2D by Regge Calculus
as an aberration of lower dimensional gravity alone
it is interesting to ask what
what Regge calculus says about the 4D transition.
This is not such a leap of faith as it might appear at first sight
- in 2D there is no classical action at all and it is the quantum
effects that contain the dynamics of the theory, whereas
in 4D there is a non-trivial classical dynamics. It is thus conceivable
that a model like RC which starts out ``close'' to smooth manifolds
might do a better job in 4D.
The Vienna group \cite{6}
have looked at
various models in 4D
including an Ising-link model
in which the edge lengths are restricted to two values
$q_l = b_l ( 1 +  \epsilon \sigma_l )$
where $\epsilon$ is a small constant
and $\sigma_l$ is an ising spin living on the link.
They found
a {\it first} order transition at positive $\beta$ (Newton coupling)
as well as 
another {\it second} order transition at negative $\beta$.

If one couples in $SU(2)$ gauge matter \cite{65}
$$S\,=\, \beta \sum_{t}A_{t}\alpha_{t} -
\frac{\beta_{g}}{2}\sum_{t}W_{t}\,{\rm Re}[ {\rm Tr}(1\,-\,U_{t})]\; 
$$ 
the gauge field string tension scaling at the negative $\beta$
transition is still what would be expected of a physical theory. 
The suggestion is then that somehow DT is looking at
the ``wrong'' first order transition - the physical theory
being the one that resides at the negative $\beta$ second
order transition.
    
\section{IN CLOSING}

In summary: the 1D results shed an interesting
light on polymer phases in higher dimensional 
models, 2D DT is in very good shape, 2D RC less so,
in 4D given the recent
results on the order of the transition  the big question is whether either the DT or RC
theories are telling us anything about Euclidean quantum gravity.
There
will no doubt be much of interest to report
at Lattice97.

Lest the reader be too despondent about the
implications of the 4D results
I would like to finish with a
couple of sentences
on recent purely analytical work   
in
Ashtekhar's approach \cite{66} to
quantum gravity
that suggest a model
not far removed from Regge Calculus or DT
emerging from a completely different perspective.
If we make $3 + 1$ split, 
and use the canonical variables
$$E_i^a E^{bi} = q q^{ab}, \; K_a^i = {1 \over \sqrt{g}} K_{ab} E^{bi}$$
where $q^{ab}$ is the 3-metric, $E^a_i$ a dreibein and $K$ the extrinsic
curvature, then a canonical transformation takes us to
Ashkekhar's ``new'' variables
$$(E^a_i, G^{-1} K_a^i ) \rightarrow ( A_a^i = G^{-1} (\gamma_a^i - i K_a^i), E_i^a)$$
where $G$ is the Newton coupling.
The constraints, a major stumbling block
in the canonical approach for many years, are greatly simplified with these variables
at the expense of a complex connection $A$.

As in lattice gauge theory
a loop representation exists 
and it was
recently realized that a spin network basis
resolves the Mandelstam constraints
that had frustrated progress in this approach.
The picture that one arrives at is, crudely, of a
embedded trivalent \footnote{A non-zero volume
actually requires higher than trivalent vertices,
but these may be broken down into trivalent parts.}
graph 
with edges coloured by spins $j$ such that
$|j_1 -j_2| \le j_3 \le j_1+j_2$.
We can even have real connection $A$ if we can put up
with more complicated constraints (i.e. Hamiltonian)
\cite{67}.
Continuum spacetime emerges from coarse graining
this lattice-like structure.
To quote Ashtekar \cite{66}:
 ``The fundamental Planck scale
excitations of the gravitational field are 1-dimensional
and the corresponding geometry is distributional'' ,
so perhaps the next presenter of this review
will not be forced to include ``on the lattice''
in the title if space(time) {\it is}  a lattice.

\end{document}